\documentclass[a4paper,11pt]{article}

\usepackage{pos}

\newcommand{\pt}{p_T}
\newcommand{\mq}{m_Q}
\newcommand{\thdc}{\theta_{\rm dc}}
\newcommand{\thLPM}{\theta_{\rm LPM}}
\newcommand{\meff}{m_{\rm eff}}
\newcommand{\QLPM}{Q_{\rm LPM}}
\newcommand{\EEC}{\mathrm{EEC}}
\newcommand{\SCETG}{\mathrm{SCET}_{\rm G}}

\title{Unraveling QCD dynamics with heavy quark energy correlators }

\author*[a]{Ivan Vitev}

\affiliation[a]{Los Alamos National Laboratory,\\
  Theoretical Division, Mail Stop B283, Los Alamos, NM 87545, USA}

\emailAdd{ivitev@lanl.gov}

\abstract{Heavy-flavor jets and their substructure provide a unique window into the role of quark mass on QCD radiation and its modification in nuclear matter. In this work, we investigate heavy-quark energy-energy  correlators (EECs) that are explicitly sensitive to mass effects, focusing on angular distributions, energy flow, and the dead-cone effect. We demonstrate how the finite mass of charm and bottom quarks reshapes the intra-jet radiation pattern, leading to characteristic suppressions at small angles and measurable deviations from massless jet expectations. Using effective field theory analysis, we extend the calculation of these observables to reactions with nuclei to show that medium-induced radiation competes with vacuum mass suppression, resulting in a nontrivial modification of jet substructure. Specifically, we identify regimes where EECs are dominated by the heavy quark mass, leading to qualitatively and quantitatively different behavior of this observable in QCD matter. We discuss the charm and bottom jet energy-energy correlators modification in reactions with nuclei, and further demonstrate how the formalism can be tested in small collision system at current facilities.}

\FullConference{The 33rd International Workshop on Deep Inelastic Scattering and Related Subjects (DIS2026)\\
4 - 8 May 2026\\
Bologna, Italy\\}

\begin{document}
\maketitle

\section{Introduction}

Energy-energy correlators (EECs) have a long history as event-shape observables in elementary collisions, where they were introduced as precision tests of perturbative QCD~\cite{Basham:1978bw}.  In modern jet physics the same idea has been reformulated as a jet-substructure observable: one measures the angular correlation of the energy flow carried by pairs of particles inside a reconstructed jet.  For a jet of transverse momentum $\pt$ and radius $R$, a convenient hadron-collider definition is
\begin{equation}
  \frac{1}{\sigma_{\rm jet}}\frac{d\Sigma}{d\theta\,d\pt\,dy}
  = \frac{1}{\sigma_{\rm jet}}
  \sum_{i,j\in {\rm jet}} \int d\sigma_{\rm jet}(\pt,y;R;\{p_k\})
  \frac{p_{T,i}p_{T,j}}{p_{T,{\rm jet}}^2}
  \delta\!\left(\cos\theta-\cos\theta_{ij}\right) .
\label{eq:eecdef}
\end{equation}
The energy weighting makes the observable insensitive to soft radiation, while its dependence on the opening angle $\theta$ gives direct access to the collinear structure of the jet.  In the small-angle regime, with $\pt\theta$ perturbative, the observable admits a factorized description and a renormalization-group evolution that can be evaluated systematically in QCD~\cite{Lee:2022uwt,Ke:2025ibt}.

Heavy-flavor jets add a new scale to this problem.  For a quark of mass $\mq$ and energy $E$, the radiation pattern changes near
\begin{equation}
  \thdc \sim \frac{\mq}{E} ,
\label{eq:deadcone}
\end{equation}
where collinear radiation is suppressed by the dead-cone effect~\cite{Dokshitzer:1991fd}.  This scale is perturbative for bottom jets over a broad range of LHC kinematics and provides a clean handle on mass effects before hadronization~\cite{Lee:2019lge} .  In QCD matter, the same heavy quark interacts with the medium through Glauber gluon exchange~\cite{Kang:2016ofv}.  The EEC then becomes sensitive not only to the vacuum mass scale but also to the medium length, screening scale, and LPM coherence scale.  Heavy-quark EECs therefore provide a differential observable with which to separate universal nuclear effects from genuinely mass-dependent jet-medium dynamics.

\section{Vacuum baseline for heavy-quark energy correlators}

The projected $N$-point energy correlator in the collinear limit can be written in terms of a hard parton production function and an energy-correlator jet function,
\begin{equation}
  \Sigma^{[N]}(R_L,\pt^2,\mq,\mu)
  = \int_0^1 dx\, x^N\,
  {J}^{[N]}(R_L,x,\mq,\mu)\,  {H}(x,\pt^2,\mu) .
\label{eq:vacfact}
\end{equation}
Here, $R_L$ is the largest angular separation in the projected correlator, $\vec H$ describes the production of the collinear source, and $\vec J^{[N]}$ contains the angular dependence and heavy-quark mass effects~\cite{Craft:2022kdo}.  For $\mu_J\sim \pt R \gg \mq$, the heavy-quark mass dependence resides in the jet function.  The nonzero-angle one-loop heavy-quark jet function has a compact form,
\begin{equation}
\begin{aligned}
 J_Q^{[N]}(R_L,\mq)\big|_{R_L\ne 0}
 &= \frac{\alpha_s C_F}{4\pi}\int_0^1 dx\,
 \frac{2\left[1-(1-x)^N-x^N\right]}
 {(-1+x)(x+\delta)(x+\bar\delta)} \\
 &\quad\times
 \left[2x^3+(1+x^2)(x+\delta)(x+\bar\delta)
 \ln\frac{\delta\bar\delta}{(x+\delta)(x+\bar\delta)}\right] .
\end{aligned}
\label{eq:heavyjet}
\end{equation}
with $\delta=i\mq/(\pt R_L)$.  At $R_L\gg \mq/\pt$ this result approaches the massless scaling regime governed by the usual twist-two anomalous dimensions.  At $R_L\sim \mq/\pt$, however, the correlator turns over and the ratio of heavy-to-light EECs reveals the suppression of small-angle radiation.  This behavior is the EEC manifestation of the dead cone and has been shown to be well described by perturbation theory for heavy-flavor jets~\cite{Craft:2022kdo}.

Precise vacuum baseline is especially important for nuclear collisions.  Because the mass dependence is perturbative, deviations from the vacuum heavy-flavor EEC can be interpreted as medium-induced modifications rather than poorly understood hadronization effects if the measured angle remains above the nonperturbative scale.

\section{Energy correlators in QCD matter}

For a jet traversing QCD matter, the factorized expression is modified by the addition of medium-induced jet functions,
\begin{equation}
\frac{d\Sigma}{d\theta\,d\pt\,dy}
= \sum_{a,b,c}\int dx_a\,dx_b\,dz_J\,   f_{a/A}(x_a,\mu) f_{b/B}(x_b,\mu)
  H_{ab\to c}\!\left(\frac{\pt}{z_J},y,\mu\right)
  \left[J^{\rm vac}_{\EEC,c}+J^{\rm med}_{\EEC,c}\right] .
\label{eq:medfact}
\end{equation}
In Eq.~(\ref{eq:medfact}) the factorization scale, jet scale, and  parton-to-jet fragmentation fraction dependence in the jet functions is implicit.
The hard scattering is unmodified at leading power, while Glauber exchanges with the medium modify the collinear evolution of the jet.  The relevant scales include
\begin{equation}
   Q\sim \pt, \qquad \mu_R\sim \pt R, \qquad \mu_\theta\sim \pt\theta,
\end{equation}
for the hard production, jet radius, and EEC measurement, together with medium-specific scales~\cite{Ke:2023ixa}
\begin{equation}
 L \; \; (\rm length), \qquad     \meff \; \; (inverse \; interaction \; range), \qquad M_{\rm LPM}\sim \sqrt{\frac{2\pt}{L}}  \; \; (LPM \; scale) .
\label{eq:mediumscales}
\end{equation}
It is further useful to define the Landau-Pomeranchuk-Migdal (LPM) angle $   \thLPM \sim \sqrt{{8\pi}/{\pt L}}$.
The region $\Lambda_{\rm QCD}/\pt \ll \theta \ll \thLPM$ is particularly interesting since in this window medium effects can enter as corrections to the anomalous dimensions that control EEC scale evolution in addition to the fixed-order contributions~\cite{Ke:2025ibt}.  The presence of a heavy quark adds the dead-cone scale $\thdc$ such that the observable is sensitive to the hierarchy between $\thdc$ and $\thLPM$.

Within $\SCETG$, the medium-induced splitting probability is organized as an opacity expansion.  At first order in opacity, the splitting rate is a sum of four contributions,
\begin{equation}
  dN^{(1),n=1}=dN^{RR}+dN^{RV}+dN^{VR}+dN^{VV},
\label{eq:rrrv}
\end{equation}
corresponding to real or virtual emissions combined with real (single-Born) or virtual (double-Born)  interactions with the medium.  The $RR$ and $RV$ terms generate the non-contact EEC at finite angle; the virtual terms are required for contact contributions and renormalization.  In a thin medium this opacity-one result captures the leading LPM physics, while higher orders provide important but typically more subtle corrections to fully differential spectra~\cite{Sievert:2019cwq}.

\section{Massive splitting functions and in-medium evolution}

The heavy-quark mass modifies the medium-induced splitting kernels in three related ways~\cite{Kang:2016ofv}.  First, it shifts the eikonal denominators through a channel-dependent mass parameter
\begin{equation}
 \nu=\begin{cases}
 (1-x)M, & Q\to Qg,\\ 
 xM, & Q\to gQ,\\
 M, & g\to Q\bar Q,
 \end{cases}
\label{eq:nu}
\end{equation}
which regulates the collinear region.  Second, it opens helicity-flipping structures absent in the massless theory,
\begin{equation}
  A_n^+ = \frac{Q_n}{Q_n^2+\nu^2}, \qquad
  A_n^- = \frac{M}{Q_n^2+\nu^2} , 
\label{eq:amplitudes}
\end{equation}
where $Q_n^2$ ate transverse momentum propagators. 
Third, it enters the LPM phases by changing the formation time,
\begin{equation}
  \Phi_n = 1-\cos\!\left[\frac{(Q_n^2+\nu^2)z^+}{2x(1-x)P^+}\right] .
\label{eq:phases}
\end{equation}
Thus, the same mass that produces the vacuum dead cone also changes the coherence pattern of medium-induced radiation. This, in turn, is reflected on the quenching pattern of heavy flavor giving an independent handle on the medium properties~\cite{Ke:2022gkq}. 

For an exponential medium profile, the path-length integral turns the oscillating LPM phase into a propagator-like factor controlled by
\begin{equation}
   Q_{\rm LPM}^2 = \frac{2x(1-x)P^+}{L^+} .
\label{eq:qlpm}
\end{equation}
This is useful since integration over variables such as the transverse momentum transfers form the QCD medium can be performed using standard Feynman parametrization.
The medium-induced heavy-quark EEC may then be written schematically as
\begin{equation}
 \left.\frac{d\Sigma}{d\theta^2}\right|_{\rm med}
 = \sum_{i\to jk}\sum_{s=\pm}\int_0^1 dx\,
 P_{ij}^{s}(x)\,R_{ij}^{s}(x,\theta;M,\QLPM,\meff),
\label{eq:heavyMedEEC}
\end{equation}
where $s=+$ and $s=-$ denote helicity-conserving and helicity-flipping contributions.  The $s=-$ term is proportional to $M^2$ and modifies the angular dependence most strongly in the dead-cone region.

\begin{figure}
    \centering
    \includegraphics[width=0.8\linewidth]{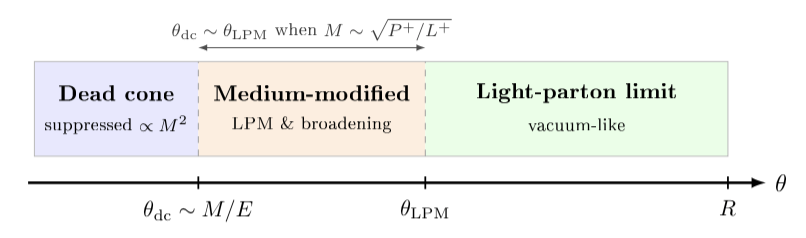}
    \caption{Schematic hierarchy of angular scales for a heavy-quark jet propagating
  through QCD matter (angle $\theta$ increasing to the right, logarithmic). At small
  angles $\theta\lesssim\theta_{\rm dc}\sim M/E$ the quark mass screens collinear
  radiation (dead cone). At intermediate angles the emission is medium-modified, with
  the coherence scale $\theta_{\rm LPM}$ (set by $\QLPM$) separating the deep-LPM
  (coherent) from the Bethe--Heitler (incoherent) regime. At large angles, up to the
  jet radius $R$, mass effects are power-suppressed and the correlator matches
  smoothly onto the light-parton limit. The dead-cone and medium-coherence scales are
  parametrically comparable, $\theta_{\rm dc}\sim\theta_{\rm LPM}$ when
  $M\sim\sqrt{P^+/L^+}$, so the two compete in the angular structure resolved by the
  EEC.}
    \label{fig:scales}
\end{figure}

The moment-space renormalization-group equation takes the form
\begin{equation}
  \mu^2\frac{d}{d\mu^2}\widetilde{\Sigma}^{[N]}_Q(\mu)
  = -\gamma_Q^{[N]}(\mu;M,\QLPM,\meff)\,
  \widetilde{\Sigma}^{[N]}_Q(\mu),
\label{eq:rge}
\end{equation}
with
\begin{equation}
  \gamma_Q^{[N]} = \gamma_{Q,{\rm vac}}^{[N]}(M,\mu)
  +\delta\gamma_{Q,{\rm med}}^{[N]}(M,\QLPM,\meff) .
\label{eq:gammadecomp}
\end{equation}
The medium correction interpolates between three regimes:
\begin{equation}
\delta\gamma_{Q,{\rm med}}^{[N]} \simeq
\begin{cases}
\delta\gamma_{\rm light}^{[N]}+{\cal O}(\nu^2/\QLPM^2), & \nu^2\ll \QLPM^2,\\
\delta\gamma_+^{[N]}(\nu^2/\QLPM^2)+\delta\gamma_-^{[N]}(\nu^2/\QLPM^2), & \nu^2\sim \QLPM^2,\\
\delta\gamma_{\rm dc}^{[N]}+\delta\gamma_-^{[N]}, & \nu^2\gg \QLPM^2.
\end{cases}
\label{eq:regimes}
\end{equation}
In the first case the mass is a perturbation and the light-parton in-medium result is recovered, but depending on kinematics mass could still play a role.  In the second case the dead-cone and LPM scales compete directly and the full massive kernel is needed.  In the third case the dead cone screens the collinear enhancement before medium coherence dominates; the remaining medium sensitivity is carried by the Coulomb logarithm and by finite helicity-flip terms. An illustration of mass and medium scales interplat is shown in Fig.~\ref{fig:scales}.

\section{Phenomenological implications}

The scale structure above implies several qualitative features for heavy flavor EECs.  In elementary collisions, the normalized two-point energy correlator for bottom and charm jets exhibits a mass-dependent suppression at small angles, with the bottom turnover occurring at larger $R_L$ than the charm turnover.  The onset of suppression in substructure scales with $M/E$, making the EEC a differential and perturbatively calculable probe of the dead cone effect~\cite{Craft:2022kdo}.  In nuclear matter, the mass-dependent turnover is modified by the competition with in-medium broadening and LPM interference.  Therefore, the heavy-quark EECs  provide a way to determine whether the medium changes only the overall energy distribution or also the scale evolution of heavy-flavor jets. The effective field theory approach also allows us to separate the universal jet dynamics from the non-perturbative properties of the medium and the theoretical framework described here is applicable to both cold (electron-nucleus collisions) and hot (heavy-ion collisions) QCD matter.
   
Our preliminary fixed-order studies indicate that different heavy-flavor producing channels are modified differently.  For a heavy quark that radiates a gluon, $Q\to Qg$, the medium can enhance the apparent dead-cone suppression at small angles.  For gluon splitting to a heavy-quark pair, $g\to Q\bar Q$, the medium can generate an enhancement near the characteristic LPM angle.  This channel dependence not only implies nontrivial structure for the  EEC modification in matter, but also points to powerful new methods that can separate modifications of an identified heavy-quark shower from those of a gluon jet that produces heavy flavor inside the jet. 

The same mass-dependent physics also appears in observables other than EECs.  For soft-drop momentum sharing distributions, earlier calculations found that in the kinematic region where mass terms dominate the splitting kernels, the modification of prompt $b$-tagged jets can be stronger than that of light jets, leading to an inversion of the usual mass hierarchy of jet-quenching effects~\cite{Li:2017wwc}.  EECs are complementary because they measure the angular distribution directly and remain closely tied to energy-flow operators and anomalous dimensions.

\begin{figure}
\centering
\includegraphics[width=0.45\linewidth]{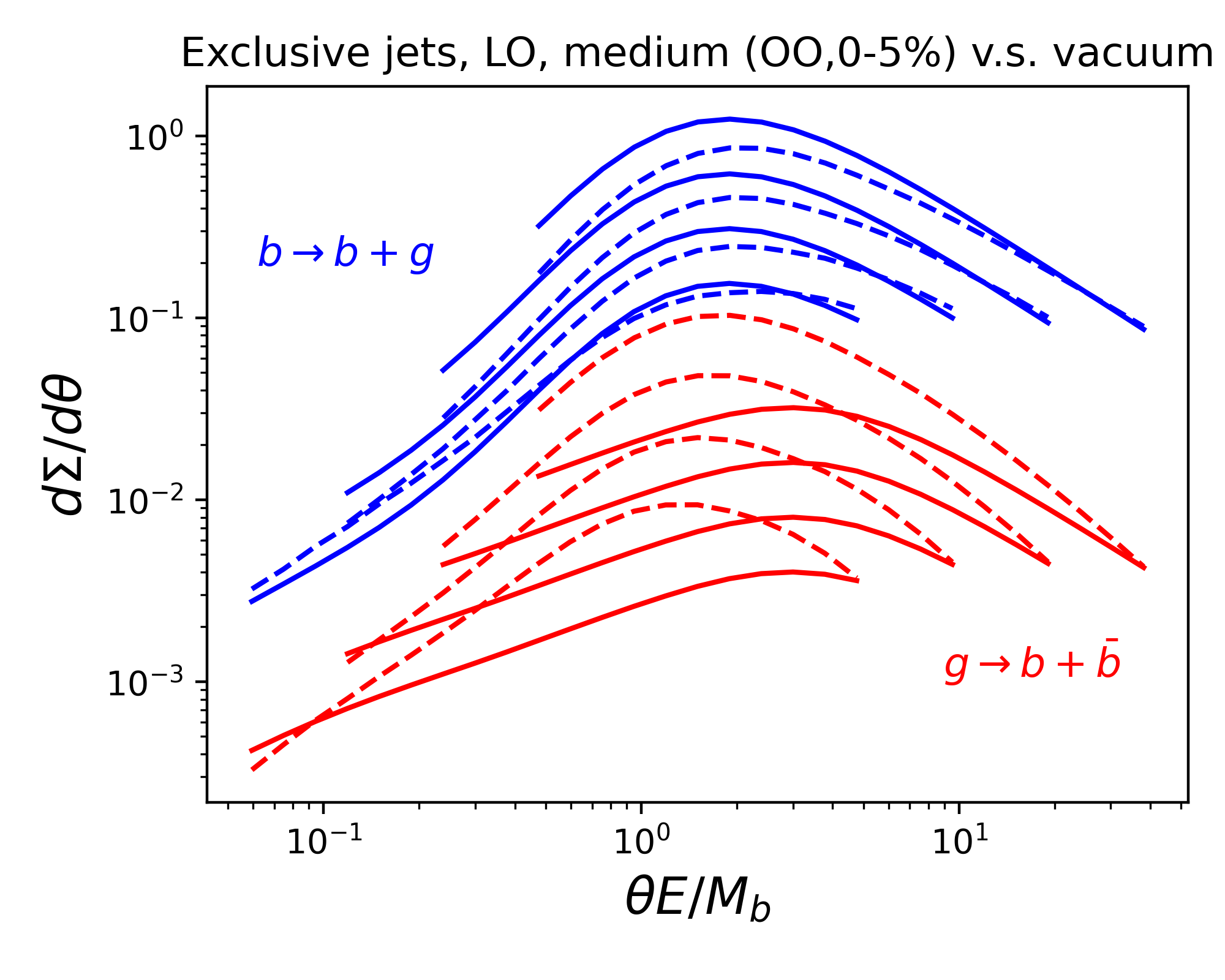}
\includegraphics[width=0.45\linewidth]{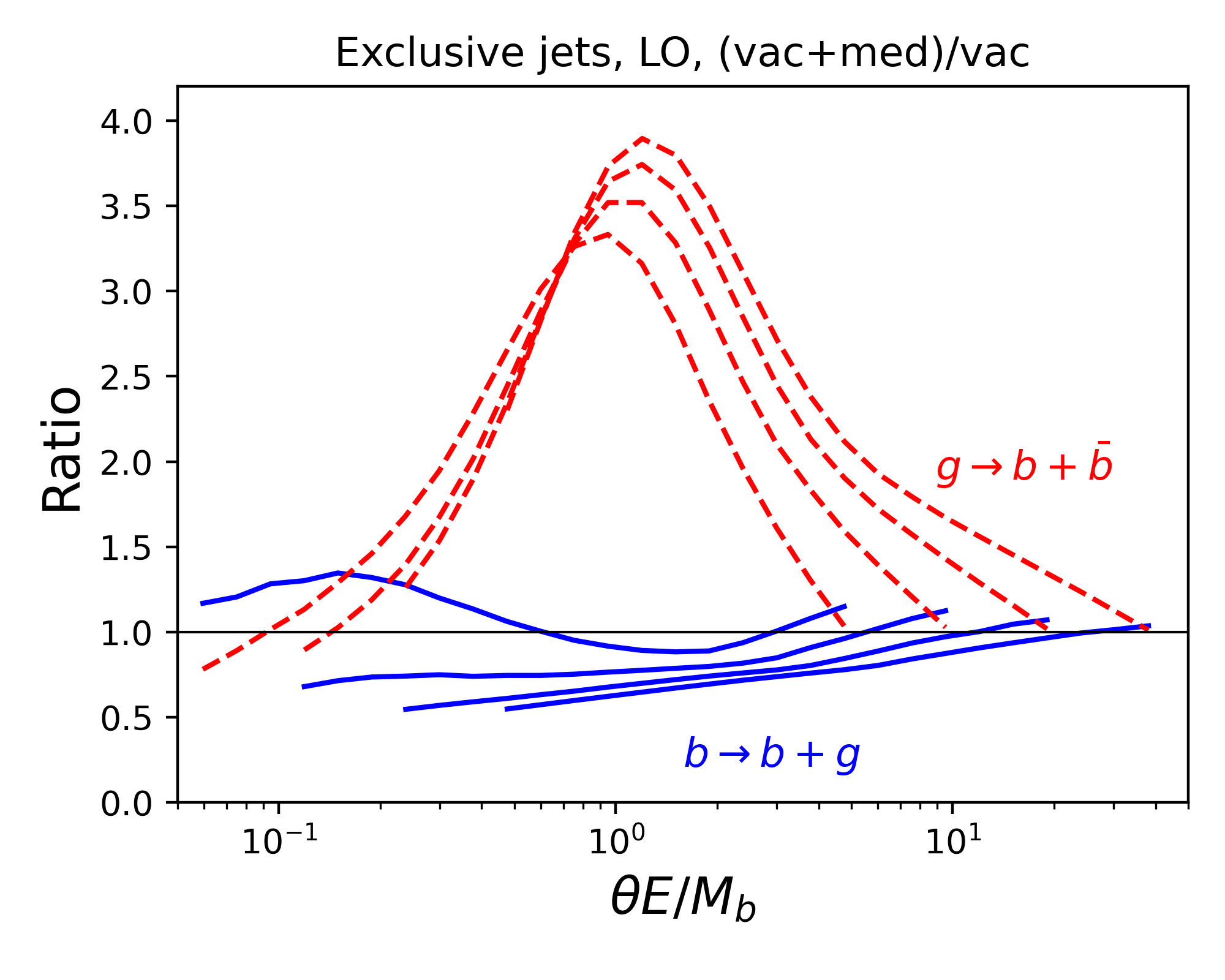}
\caption{Left: exclusive jet function for $b$-jet EEC in the medium (dashed) and the vacuum (solid) at different jet energies. The red lines are for $g\rightarrow b+\bar{b}$ channel and the blue solid lines are for $b\rightarrow b+g$ channel. Right: ratio of exclusive jet function in the medium to those in the vacuum. The medium profile is obtained from hydrodynamic simulation for O+O in 0-5\% centrality and the jet function is numerically integrated in the hydrodynamic medium. As seen in the ratio plot, both the $b\rightarrow b+g$ and the $g\rightarrow b+\bar{b}$ channels are sensitive to the presence of a medium but in different ways. For the former, as the jet energy is reduced, the additional suppression can transition to enhancement at small angles.}
\label{fig:Non-contact-b-jet-in-medium-ratio}
\end{figure}

Experimentally, charm- and bottom-tagged EECs can be studied at the LHC, RHIC (from data on tape), and in future EIC measurements.  The most robust region is expected to be above the nonperturbative scale but below or near the LPM angle, where the theory predicts a controlled medium correction to the evolution.  At large angles of order the jet radius, energy loss and medium response can contribute significantly; at very small angles, hadronization and finite tracking thresholds may become important.  A systematic comparison of charm, bottom, and light-jet EECs across $p$--Pb, O--O, and Pb--Pb systems would test how the medium-induced scale evolution depends on the system size and whether heavy-flavor EECs can constrain transport properties and path-length dependence of nuclear matter. The heavy-ion systems available today can further validate the in-medium EEC theory for future applications to deep inelastic scattering on nuclei at the EIC. Preliminary results for the EEC jet functions only are shown in Fig.~\ref{fig:Non-contact-b-jet-in-medium-ratio}, see caption for details.

\section{Conclusions}

Heavy-quark energy correlators combine the perturbative control of energy-flow observables with the intrinsic mass scale of charm and bottom quarks.  In vacuum, the mass introduces a calculable dead-cone suppression of the EEC below the characteristic  $\theta\sim M/E$.  In QCD matter, this scale competes with Glauber-induced broadening and the LPM effect.  The resulting in-medium heavy-quark EEC is governed by massive opacity-one splitting functions, with a mass-regulated helicity-conserving piece and a genuinely new helicity-flipping contribution.  The corresponding anomalous dimensions interpolate smoothly between the light-parton limit and a dead-cone-dominated regime. Preliminary phenomenological results are very encouraging and show that the $Q\to Qg$ and $g\to Q\bar Q$ channels can be modified in qualitatively different ways. Charm and bottom quark jets exhibit distinct angular turnoffs and comparisons across collision systems can isolate medium length, screening, and broadening effects.  Heavy-flavor EEC measurements could, therefore, complement traditional jet-shape, fragmentation, and soft-drop observables, and could provide a precision path toward extracting mass-dependent jet-medium dynamics from QCD. In the future such work can be extended to  azimuthal-anisotropic substructure~\cite{Ke:2024emw} and EECs with quarkonia~\cite{Copeland:2025osx,Copeland:2026yqa}.  

\section*{Acknowledgments}

The author thanks Weiyao Ke and Bianka Me\c{c}aj for collaboration on the work summarized here.  The work of B.M. and I.V. is supported by the U.S. Department of Energy through Los Alamos National Laboratory.  Los Alamos National Laboratory is operated by Triad National Security, LLC, for the National Nuclear Security Administration of the U.S. Department of Energy under Contract No. 89233218CNA000001.  The research performed by W.K., B.M., and I.V. is further funded by LANL's Laboratory Directed Research and Development program.

\end{document}